\documentclass[aps,pra,twocolumn,showpacs,preprintnumbers,amsmath,amssymb,footinbib,floatfix]{revtex4}
\usepackage{graphicx,epsfig}
\usepackage{bm}
\usepackage{dcolumn}
\usepackage{xcolor}
\usepackage[breaklinks=true,colorlinks,citecolor=blue,linkcolor=blue,urlcolor=blue]{hyperref}
\usepackage{amsmath}
\usepackage{lipsum}
\usepackage{upgreek}
\usepackage{footnote}
\usepackage{natbib}

\begin{document}

\title{A method to remove lower order contributions in multi-particle femtoscopic correlation functions
}


\author{Raffaele Del Grande$^1$}\email{raffaele.del-grande@tum.de}        \author{Laura Šerkšnytė$^1$}\email{laura.serksnyte@tum.de}
\author{Laura Fabbietti$^1$}
\author{Valentina~Mantovani~Sarti$^1$} 
\author{Dimitar Mihaylov$^1$}

\affiliation{$^1$Physik Department E62, Technische Universit{\"a}t M{\"u}nchen, Garching, Germany}


\begin{abstract}
In recent years the femtoscopy technique has been used by the ALICE Collaboration in small colliding systems at the LHC to investigate the strong-interaction of hadron pairs in the low-energy regime. 
The extension of this technique to the study of many-body correlations aims to deliver in the next years the first experimental measurements of the genuine many-hadron interactions, provided that the contributions due to the lower order terms are properly accounted for. In this paper we present a method that allows to determine the residual lower order contributions to the three-body correlation functions, based on the cumulant decomposition approach and on kinematic transformations. A procedure to simulate genuine three-body correlations in three-baryon correlation functions is also developed. 
A qualitative study of the produced correlation signal is performed by varying the strength of the adopted three-body interaction model and comparisons with the expectations for the lower order contributions to the correlation function 
are shown. 
The method can be also applied to evaluate the combinatorial background in the two-body correlation functions, providing an improved statistical accuracy with respect to the standard techniques.
The example of the contribution by the pK$^+$K$^-$ channel to the recently measured p$\upphi$ correlation is discussed. 
\end{abstract}
\maketitle

\section{Introduction} \label{s:intro}
One of the main challenges of the modern nuclear physics is to obtain a quantitative description of the many-body strong interaction among hadrons. 
The many-body dynamics is crucial in the theoretical explanation of various phenomena, such as the absorption processes of mesons in nuclei \cite{FriedmanGal} and the observations of heavy neutron stars (M $>$ 2 M$_\odot$) \cite{Lonardoni:2014bwa}, which are apparently in contradiction with the formation of hyperons in the interior of such stars. 
Many-body scattering and bound states calculations involving hyperons and nucleons are typically performed employing effective interaction potentials \cite{faddeev2016scattering,Yakubovsky:1966ue}. Alternative approaches to the many-body problem make use of variational methods, such as the so-called Green's function Monte Carlo and the quantum Monte Carlo techniques \cite{pudliner1997quantum,WIRINGA1984273,Breunig2000}. So far, experimental inputs to the three-body interaction models in the strangeness $S = 0$ and $S=1$ sectors are provided by the measurements of nuclei and hypernuclei binding energies, not ideal to constrain many-body scattering calculations.

In the last two decades the femtoscopy method \cite{wiedemann1999particle,heinz1999two,lednicky2004correlation,lisa2005femtoscopy} was applied in collider and fixed target experiments at energies varying from few GeV to few TeV. In particular, the hadrons which are produced in pp and p--Pb collisions at the LHC are emitted at small relative distances (of the order of 1 fm) and may undergo Final State Interactions (FSIs) before the detection. The interacting hadrons are correlated in the momentum space and the underlying dynamics can be tested by studying the correlation function \cite{lednicky2004correlation,lisa2005femtoscopy}. 
Recently, the method was applied by the ALICE Collaboration and was used to study FSIs of hadrons produced in pp and p--Pb collisions. 
The high statistics collected by ALICE allowed to precisely measure the correlation function for multiple hadron pairs (p-p \cite{acharya2019p}, pK$^+$ and pK$^-$ \cite{Acharya:2019bsa}, p$\mathrm{\Lambda}$ \cite{acharya2019p}, p$\mathrm{\Sigma}^0$ \cite{acharya2020investigation}, $\mathrm{\Lambda}\mathrm{\Lambda}$ \cite{acharya2019study}, p$\mathrm{\Xi}^-$ \cite{acharya2019first}, p$\mathrm{\Omega}^-$ \cite{nature}, p$\upphi$ \cite{alice2021experimental} and $\mathrm{B\bar{B}}$ \cite{Acharya_2020}) and several models for the two-body strong interaction could be validated (for a complete review see Ref. \cite{fabbietti2020hadronhadron}). 

A first extension of this technique in the many-particle sector can be obtained by addressing three-body correlations \cite{Heinz_2004,Gangadharan_2015}, a measurement which will become feasible within the next years. One of the crucial aspects in the interpretation of the three-body correlations is represented by the influence of the lower order, two-body, interactions on the collected triplets sample. A first study with pion-triplets measured in Ref. \cite{PhysRevC.89.024911} demonstrated that the Kubo's cumulant expansion method \cite{kubo1962generalized} can be employed to isolate and subtract the contributions of two-body quantum statistics (QS) from the three-body systems.
This contribution was obtained using a data-driven approach, by selecting two-pions produced in the same collision and a third pion from a different event to avoid three-body correlations. 
The same procedure can be applied to a sample of any three hadrons to isolate the genuine three-body correlation due to the strong interaction. 
Alternatively, the lower order contributions can be calculated starting from the genuine two-body correlation functions, either the measured or the theoretical, imposing the kinematic constraints of the analysed three-body system. The advantage of this method is that it is 
not biased by the experimental procedure of mixing particles emitted in the same collision with particles emitted in a different collision. 
In this paper, the latter procedure, called the projector method, will be presented. Toy Monte Carlo (MC) simulations can be employed to study the shape of three-body correlation functions, including genuine two- and three-body contributions. To this end, models for the two- and three-body interactions are assumed. Using the projector method to set the baseline due to the lower order correlations, it is possible to qualitatively study the simulated correlation function varying the strength of the three-body forces.

Furthermore, the method allows the projection of correlation functions in any kinematic variable of interest. This renders the projector method formalism useful also for the evaluation of other sources of background which affect the measurement of genuine hadron-hadron correlation functions. An example is the combinatorial background in the identification of unstable hadrons from their decay daughters. The latter are experimentally contaminated by particles of different origin, entering as a background in the measured correlation function.  
In the case of p$\mathrm{\upphi}$ analysis the correlation function is affected by the residual pK$^+$, pK$^-$ and K$^+$K$^-$ correlations (see Ref. \cite{alice2021experimental}). 
The problem is commonly solved by constructing the correlation function starting from particles in the side-bands region of the invariant mass of the particle of interest. In the specific example of the $\mathrm{\upphi}$ meson, one would consider the side-bands on the K$^+$K$^-$ invariant mass. The obtained correlation can be used as a baseline for the interaction studies \cite{acharya2020investigation}.
The method developed in this work will allow to evaluate theoretically the projection of the pK$^+$, pK$^-$ and K$^+$K$^-$ correlations in the p$\mathrm{\upphi}$ correlation function, leading to an improved precision of the background model.

The paper is organised as follows:  
general concepts on the two- and three-body correlation functions are given in Sections \ref{S2body} and \ref{S3body}; the method is formalised and developed in Section \ref{SectionProjector} and tests are performed in Section \ref{SMC} by means of toy MC simulations of a two-body correlation in three-body systems; in Section \ref{Sppp-ppp} genuine three-body correlations are introduced in the simulated data sample and a study of the generated correlation function is performed; the evaluation of the pK$^+$K$^-$ combinatorial background contribution in the p$\upphi$ correlation function is calculated in Section \ref{Spphi} and the conclusions are given in Section \ref{s:conclusion}.

\section{Two-body correlation function}\label{S2body}
The femtoscopic two-body correlation function is defined in terms of the relative momentum $\mathbf{k}$ of the interacting pair and is calculated using the Koonin-Pratt formula \cite{koonin1977proton,pratt1990detailed}
\begin{equation}
    C_2(\mathbf{k}) =  \int_{V} S_2(\mathbf{r}) \ |\psi_{\mathbf{k}} (\mathbf{r})|^2 \  d^3\mathbf{r} \ ,
    \label{KPformula}
\end{equation}
where $S_2(\mathbf{r})$ is the two-body source function, which is normalised to the unity in the domain $V$ of the relative position coordinates $\mathbf{r}$, $\psi_{\mathbf{k}} (\mathbf{r})$ is the wave function of the pair. 
In the non-relativistic regime, the wave function $\psi_{\mathbf{k}}(\mathbf{r})$ is a solution of the stationary Schr{\"o}dinger equation, given by the Hamiltonian operator of the relative system
\begin{equation}
     \mathcal{H}_{r} =  \frac{\mathbf{k}^2}{2 \mu} + V_2(\mathbf{r})  =  - \frac{\hbar^2 }{2 \mu} \mathbf{\nabla}^2_\mathbf{r} + V_2(\mathbf{r}) \ ,
\end{equation}
where $\mu$ is the reduced mass of the system and $V_2(\mathbf{r})$ is the two-body interaction potential. Under the hypotheses of no explicit time and momentum dependence for the particles emitter, a Gaussian two-body source function can be assumed 
\begin{equation}
    S_2(\mathbf{r}) = \frac{1}{(4 \pi r_0^2)^{3/2}} \ e^{- \frac{\mathbf{r}^2}{4 r_0^2}} \ ,
    \label{2BS}
\end{equation}
where $r_0$ is the emitting source radius (more details on the source function are discussed in Refs. \cite{lisa2005femtoscopy,fabbietti2020hadronhadron,Mihaylov:2018rva}).

Given two particles with masses $m_1$ and $m_2$, position coordinates $\mathbf{x}_1$ and $\mathbf{x}_2$, the Koonin-Pratt formula can be generalised, including the center of mass (CM), as
\begin{equation}
\resizebox{0.92\hsize}{!}{
     $
    C_2(\mathbf{p}_1, \mathbf{p}_2) = \int_{V_1} \int_{V_2} S_2(\mathbf{x}_1, \mathbf{x}_2) \ |\psi_{\mathbf{p}_1, \mathbf{p}_2}(\mathbf{x}_1, \mathbf{x}_2)|^2 \ d^3\mathbf{x}_1 d^3 \mathbf{x}_2 \ ,
    $
    }
    \label{2bCF}
\end{equation}
where $\mathbf{p}_1$ and $\mathbf{p}_2$ are the particle momenta and $\psi_{\mathbf{p}_1, \mathbf{p}_2}(\mathbf{x}_1, \mathbf{x}_2)$ is now the wave function of the two-body system in the position coordinates representation. The non-relativistic Hamiltonian operator turns to be 
\begin{equation}
    \mathcal{H} = \frac{\mathbf{p}_1^2}{2 m_1} + \frac{\mathbf{p}_2^2}{2 m_2} + V_2(\mathbf{x}_1 - \mathbf{x}_2) \ .
\end{equation}
Applying the transformation from the Cartesian to the relative coordinates basis in Eq. \eqref{2bCF}, the Koonin-Pratt formula in Eq. \eqref{KPformula} is recovered. In the absence of interactions, i.e. $V_2(\mathbf{x}_1 - \mathbf{x}_2) = 0$, the wave function of the system is given by the product of plane waves and the correlation function becomes $C_2(\mathbf{p}_1, \mathbf{p}_2) = 1$.

\section{Three-body correlation function}\label{S3body}
The definitions presented in Section \ref{S2body} are now extended to a system of three particles having masses $m_1$, $m_2$ and $m_3$. Starting from the general Eq. \eqref{2bCF}, the three-body correlation function can be defined as
\begin{equation}
\begin{aligned}
    C_3(\mathbf{p}_1, \mathbf{p}_2, \mathbf{p}_3) = \int_{V_1} \int_{V_2} \int_{V_3} S_3(\mathbf{x}_1, \mathbf{x}_2, \mathbf{x}_3) \times \\ \times |\psi_{\mathbf{p}_1, \mathbf{p}_2, \mathbf{p}_3}(\mathbf{x}_1, \mathbf{x}_2, \mathbf{x}_3)|^2 \ d^3\mathbf{x}_1 d^3 \mathbf{x}_2 d^3 \mathbf{x}_3 \ ,
\end{aligned}
   \label{3bCF}
\end{equation}
where $\mathbf{x}_1$, $\mathbf{x}_2$ and $\mathbf{x}_3$ are the coordinates of the three particles, $\mathbf{p}_1$, $\mathbf{p}_2$ and $\mathbf{p}_3$ are the corresponding conjugate momenta, $S_3(\mathbf{x}_1, \mathbf{x}_2, \mathbf{x}_3)$ is the three-body source function, which is normalised to the unity in the 9-dimensional $V_1 \times V_2 \times V_3$ volume and $\psi_{\mathbf{p}_1, \mathbf{p}_2, \mathbf{p}_3}(\mathbf{x}_1, \mathbf{x}_2, \mathbf{x}_3)$ is the wave function of the three-body system in the position coordinates representation. 
The three-body correlation function is given as a function of the Lorentz invariant hyper-momentum $Q_3$ \cite{PhysRevC.89.024911,Heinz_2004} defined as
\begin{equation}
    Q_3 = \sqrt{-q_{12}^{2} - q_{23}^2 - q_{31}^2} \ ,
    \label{Q3general}
\end{equation}
where $q_{ij}$ is the modulus of the four-vector
\begin{equation}
    q_{ij}^\mu = \frac{2\ m_j}{m_i + m_j}\ p_{i}^\mu - \frac{2\ m_i}{m_i + m_j}\  p_{j}^\mu \ .
\end{equation}
In the non-relativistic approximation 
Eq. \eqref{Q3general} reads
\begin{equation}
     Q_3 = 2\ \sqrt{\sum_{i<j = 1}^{3} \left(\frac{m_j}{m_i+m_j} \mathbf{p}_i - \frac{m_i}{m_i+m_j} \mathbf{p}_j \right)^2 }.
     \label{Q3cart}
\end{equation}
For each given value of $Q_3$, multiple configurations in the momentum phase space $(\mathbf{p}_1, \mathbf{p}_2, \mathbf{p}_3) \in \mathcal{S}$ are allowed. 
The three-body correlation function can be calculated as a function of $Q_3$ by performing the integration over all the momentum configurations corresponding to the same value of $Q_3$, i.e. 
\begin{equation}
\resizebox{0.9\hsize}{!}{
     $
    C_3(Q_3) = \iiint_{(\mathbf{p}_1, \mathbf{p}_2, \mathbf{p}_3) \in \mathcal{D}_{3}} C_3(\mathbf{p}_1, \mathbf{p}_2, \mathbf{p}_3)\ \mathcal{N} \ d^3\mathbf{p}_1 \ d^3\mathbf{p}_2\  d^3\mathbf{p}_3 \ ,$}
    \label{masterProj}
\end{equation}
where $\mathcal{D}_{3}$ is the integration domain defined as follows
\begin{equation}
    \mathcal{D}_{3} = \{ (\mathbf{p}_1, \mathbf{p}_2, \mathbf{p}_3) \in \mathcal{S}\ |\ Q_3 = \mathrm{constant} \} \ .
    \label{domain}
\end{equation}
The density of states for each configuration is uniform, 
thus the normalization constant $\mathcal{N}$ in Eq. \eqref{masterProj} is defined as
\begin{equation}
    \mathcal{N} = \left[ \iiint_{(\mathbf{p}_1, \mathbf{p}_2, \mathbf{p}_3) \in \mathcal{D}} \ d^3\mathbf{p}_1 \ d^3\mathbf{p}_2\  d^3\mathbf{p}_3 \right]^{-1} \ .
\end{equation}
In the absence of interaction the three-body correlation function is equal to the unity $C_3(\mathbf{p}_1, \mathbf{p}_2, \mathbf{p}_3) = 1$ and the projection onto the $Q_3$ scalar is $C_3(Q_3) = 1$.

The formula in Eq. \eqref{masterProj} can be generalised to the $N$-body system given the kinematic variable of interest, denoted here as $Q_N$
\begin{equation}
\resizebox{0.9\hsize}{!}{
     $
    C_N(Q_N) = \int \cdots \int_{(\mathbf{p}_1, ... ,  \mathbf{p}_N) \in \mathcal{D}_N} C_N(\mathbf{p}_1, ..., \mathbf{p}_N)\ \mathcal{N} \ d^3\mathbf{p}_1 \ \cdots d^3\mathbf{p}_N \ ,$}
    \label{masterProjN}
\end{equation}
where the domain $\mathcal{D}_N$ and the normalisation $\mathcal{N}$ are defined accordingly.

\section{Two-body contributions in three-body correlation functions}
\label{SectionProjector}
In three-body correlation studies, three particles emitted in the same collision are measured. The correlation in the particles momentum space is provided both by the pairwise interactions among the hadrons and, eventually, by the three-body interaction. The contribution due to the genuine three-body correlation can be isolated using the Kubo's cumulant expansion method (see Ref. \cite{kubo1962generalized} for the details).  
Using the Kubo's rule, a triplet sample denoted with $(1,2,3)$, where the numbers indicate the particles, can be decomposed as follows
\begin{equation}
\begin{aligned}
    (1,2,3) &= -\ 2\ \times \ (1)(2)(3) \ +\\
    &+\ ([1,2],3)\ +  \ ([2,3],1)\ +\ ([3,1],2)\ +\\
    &+\ ([1,2,3]) \ ,
\end{aligned}
\label{eq:KuboDec}
\end{equation}
where $([i,j],k)$ are statistical sub-samples in which only the particles in the square brackets are correlated; $(i)(j)(k)$ is the sub-sample of three uncorrelated particles and $([i,j,k])$ is the genuine three-body correlated sample. 
The correlation function corresponding to Eq. \eqref{eq:KuboDec} is
\begin{equation}
\begin{aligned}
    C_3(\mathbf{p}_1, \mathbf{p}_2, \mathbf{p}_3) &= C_3 ([\mathbf{p}_1, \mathbf{p}_2],\mathbf{p}_3) + C_3 (\mathbf{p}_1, [\mathbf{p}_2,\mathbf{p}_3]) +\\
    &+ C_3 (\mathbf{p}_2, [\mathbf{p}_3,\mathbf{p}_1]) -\ 2\ +\\ &+ \mathbf{c}_3 ([\mathbf{p}_1, \mathbf{p}_2,\mathbf{p}_3]) \ ,
\end{aligned}
\label{cumulantsCF}
\end{equation}
where $\mathbf{c}_3 ([\mathbf{p}_1, \mathbf{p}_2,\mathbf{p}_3])$ is the three-body cumulant (see Ref. \cite{PhysRevC.89.024911} for the formal derivation). 
In the absence of a genuine three-body interaction the cumulant is equal to zero ($\mathbf{c}_3 = 0$) and the three-body correlation function is 
\begin{equation}
\begin{aligned}
    C_3(\mathbf{p}_1, \mathbf{p}_2, \mathbf{p}_3) &= C_3 ([\mathbf{p}_1, \mathbf{p}_2],\mathbf{p}_3) + C_3 (\mathbf{p}_1, [\mathbf{p}_2,\mathbf{p}_3]) +\\ &+ C_3 (\mathbf{p}_2, [\mathbf{p}_3,\mathbf{p}_1]) - 2 \\
    &\equiv C_3^{2b}(\mathbf{p}_1, \mathbf{p}_2, \mathbf{p}_3)  \ ,
    \label{Cum}
\end{aligned}
\end{equation}
where the superscript $2b$ indicates that only the two-body interactions are considered.

Given that in each statistical sub-sample $([i,j],k)$ only the pair $(i,j)$ interacts, 
the three-body correlation functions $C_3 ([\mathbf{p}_i, \mathbf{p}_j],\mathbf{p}_k)$ in Eq. \eqref{Cum} can be calculated by using the two-body correlation function of the $(i,j)$ pair.  
In this specific case the Hamiltonian of the system is 
\begin{equation}
    \mathcal{H} = \mathcal{H}_{CM} + \mathcal{H}_{1} + \mathcal{H}_{2}  = \frac{\mathbf{P}^2}{2\ M} + \left[  \frac{\mathbf{k}_1^2}{2 \ \mu_1} + V(\mathbf{r}_1) \right] + \frac{\mathbf{k}_2^2}{2 \ \mu_2} \ ,
    \label{H3rel}
\end{equation}
where $\mathbf{P}$ is the CM momentum, $\mathbf{k}_1$ and $\mathbf{k}_2$ are the relative momenta for the three-body system corresponding to the Jacobi coordinates $\mathbf{r}_1$ and $\mathbf{r}_2$, $M$ is the total mass, $\mu_1$ and $\mu_2$ are the reduced masses (the definitions are given in Appendix \ref{formalisation}).  
Since the Hamiltonian operators $\mathcal{H}_{CM}$, $\mathcal{H}_{1}$ and $\mathcal{H}_{2}$ commute, the total wave function of the system 
can be factorised. The solutions of the stationary Schr{\"o}dinger equations for the Hamiltonians $\mathcal{H}_{CM}$ and $\mathcal{H}_{2}$ are free plane waves in the position coordinates representation and then the three-body correlation function defined in Eq. \eqref{3bCF} turns to be
\begin{equation}
\resizebox{0.95\hsize}{!}{
     $
       C_3(\mathbf{P}, \mathbf{k}_1, \mathbf{k}_2) =  \int_{V_{\mathbf{R}}} \int_{V_{\mathbf{r}_1}} \int_{V_{\mathbf{r}_2}} S_3(\mathbf{R}, \mathbf{r}_1, \mathbf{r}_2) \ |\psi_{1}(\mathbf{r}_1 )|^2 \ d^3\mathbf{R}\ d^3\mathbf{r}_1 d^3 \mathbf{r}_2 \ ,
       $
       }
       \label{C3b}
\end{equation}
where the wave function $\psi_{1}(\mathbf{r}_1)$ is a solution of the Schr{\"o}dinger equation for the Hamiltonian operator $\mathcal{H}_1$ in Eq. \eqref{H3rel}.
Given that the wave function in Eq. \eqref{C3b} depends only on the coordinate $\mathbf{r}_1$, the source function can be marginalised in $\mathbf{r}_1$ by performing the integrals in $d^3\mathbf{R}$ and $d^3\mathbf{r}_2$, i.e.
\begin{equation}
    S_3(\mathbf{r}_1) = \int_{V_{\mathbf{R}}} \int_{V_{\mathbf{r}_2}} S_3(\mathbf{R}, \mathbf{r}_1, \mathbf{r}_2) \ d^3\mathbf{R}\ d^3\mathbf{r}_2 \ .
    \label{3BS}
\end{equation}
Assuming that the two- and three-body sources have the same radius, evaluated in Ref. \cite{2020135849}, 
we get the equivalence of the three-body and two-body source functions $S_3(\mathbf{r}_1) = S_2(\mathbf{r}_1)$ 
and, consequently, the equivalence of the three-body and the two-body correlation functions, recovering the Koonin-Pratt formula in Eq. \eqref{KPformula}, i.e. $C_3(\mathbf{P}, \mathbf{k}_1, \mathbf{k}_2) = C_2 (\mathbf{k}_1)$.

Following the definitions given is Section \ref{S3body}, the correlation function is calculated as a function of the scalar $Q_3$. 
To this end Eq. \eqref{Q3cart} is rewritten in terms of the relative momenta 
\begin{equation}
Q_3 =  \sqrt{ \alpha \ \mathbf{k}_1^2 + 2 \beta\ \mathbf{k}_1 \cdot \mathbf{k}_2 + \gamma\ \mathbf{k}_2^2} 
     \label{Q3rel}
\end{equation}
where the constants $\alpha$, $\beta$ and $\gamma$ depend on the particles mass\footnote{ $\alpha = \frac{4\ m_3^2}{(m_1+m_3)^2}+\frac{4\ m_3^2}{(m_2+m_3)^2}+ 4 \ ;$\\ $\beta = \frac{4\ m_3 (m_1+m_2+m_3)}{m_1+m_2}
   \left[ \frac{m_2}{(m_2+m_3)^2}-\frac{m_1}{(m_1+m_3)^2}\right] \ ;$\\ $\gamma  = \frac{4\ (m_1+m_2+m_3)^2}{(m_1 + m_2)^2}
   \left[\frac{ m_1^2}{(m_1 + m_3 )^2}+\frac{m_2^2}{( m_2 + m_3)^2}\right] \ .$}.

The Eq. \eqref{Q3rel} for a fixed value of $Q_3$ is the analytical formula of a rotated hyper-ellipsoid in ${\rm I\!R}^{6}$, then using a parametrisation for the surface, the integral in Eq. \eqref{masterProj} becomes
\begin{equation}
\resizebox{0.99\hsize}{!}{
$
\begin{aligned}
    &C_3(Q_3) = \iiiint_{Q_3 = \text{const}} C_2({k}_1) \ \mathcal{N} \ k_1^2 \ k_2^2 \ d{\Omega}_1 d{\Omega}_2 d{k}_1 d{k}_2 = \\
    &= \int_0^{\sqrt{\frac{\gamma}{\alpha \gamma - \beta^2}} \ Q_3} C_2(k_1) \ \left[ \frac{16 (\alpha \gamma - \beta^2)^{3/2} k_1^2}{\pi Q_3^4 \gamma^2} \sqrt{\gamma Q_3^2 - (\alpha \gamma - \beta^2) k_1^2}\right] \  dk_1 \ .
\end{aligned}
$}
    \label{PQ3}
\end{equation}
The Eq. \eqref{PQ3} projects the two-body correlation function $C_2(k_1)$ 
from the two-body relative momentum $k_1$ on the three-body hyper-momentum $Q_3$. The projection is provided by the following analytical function 
\begin{eqnarray}
    W(k_1,Q_3) = \frac{\iiint k_1^2 k_2^2 dk_2 d\Omega_1 d\Omega_2}{\iiiint k_1^2 k_2^2 dk_1 dk_2 d\Omega_1 d\Omega_2} = \\
    = \frac{16 (\alpha \gamma - \beta^2)^{3/2} k_1^2}{\pi Q_3^4 \gamma^2} \sqrt{\gamma Q_3^2 - (\alpha \gamma - \beta^2) k_1^2} 
    \label{projector}
\end{eqnarray}
that represents the density of states in the phase space in the volume $(k_1, k_1 + dk_1)$ for a fixed value of $Q_3$. The same procedure can be applied to calculate the two-body correlation functions in any chosen kinematic variable by writing the integration domain in Eq. \eqref{domain}  accordingly. An example will be shown in Section \ref{Spphi}.

Performing the projection onto $Q_3$ for all the terms in the right side of Eq. \eqref{Cum}, the total contribution to the three-body correlation function due to the two-body interactions is obtained as a function of $Q_3$
\begin{equation}
    C_3^{2b} (Q_3) = C_3^{12} (Q_3) + C_3^{23} (Q_3) + C_3^{31} (Q_3) - 2
    \label{sumC3}
\end{equation}
where the indices $ij$ refer to the label of the interacting particle pairs. Each term of the sum is calculated using the two-body correlation function for the interacting pairs, i.e.
\begin{equation}
    C_3^{ij}(Q_3) = \int C_2(k_1^{ij}) \ W^{ij}(k_1^{ij},Q_3) \ dk_1^{ij} \ ,
    \label{kprojQ3}
\end{equation}
where the function $W^{ij}$ is the projector defined in Eq. \eqref{projector}.

\section{Validation of the method}\label{SMC}
The projector method developed in Section \ref{SectionProjector} is tested by means of a toy Monte Carlo (MC) simulation. A data sample of three particles with masses $m_1$, $m_2$ and $m_3$ and momenta $\mathbf{p}_1$, $\mathbf{p}_2$, $\mathbf{p}_3$ is generated, assuming that only the pair (1,2) is correlated in the momentum space.  
Given the vector components of the particle momenta, the relative momentum of the correlated pair $k_{1}$ and the scalar $Q_{3}$ are calculated event-by-event. Since $k_{1}$ is Lorentz invariant, we will use from now on the common notation $k^{*}$ which is the relative momentum of the interacting pair in their CM frame. The two-body correlation function for the pair (1,2) is then obtained as a function of $k^{*}$ using the formula used in femtoscopy to calculate the experimental correlation function \cite{lisa2005femtoscopy}, i.e.
\begin{equation}
    C_2(k^{*}) = \frac{N_\mathrm{same}(k^{*})}{N_\mathrm{mixed}(k^{*})} \ ,
    \label{eq:C2PM}
\end{equation}
where $N_{\rm{same}}(k^{*})$ is the relative momentum distribution of pairs of particles produced in the same MC event and $N_{\rm{mixed}}(k^{*})$ is obtained  using the so-called mixed event technique, combining particles generated in different MC events. 

The three-body correlation function of the triplet is also calculated.  In this case, the same and mixed event distributions as a function of $Q_3$ are used, i.e. 
\begin{equation}
    C_3(Q_3) = \frac{N_\mathrm{same}(Q_3)}{N_\mathrm{mixed}(Q_3)} \ .
    \label{eq:C3}
\end{equation}
The correlation function in Eq. \eqref{eq:C3} is finally compared to the prediction of the projector method, using the two-body correlation function $C_2(k^{*})$ as input in Eq. \eqref{kprojQ3}. 

\subsection{MC data sample}\label{sec:datasample}
The simulated data sample is generated following the Kubo's decomposition rule in Eq. \eqref{eq:KuboDec}, i.e. adding correlated and uncorrelated particle sub-samples.
The first hypothesis within the simulation is that only the pair (1,2) is correlated, consequently the (2,3) and (3,1) pairs correlations and the three-body cumulant are assumed to be zero. The first two terms in Eq. \eqref{eq:KuboDec} are then considered. Such assumptions will be removed in Section \ref{Sppp-ppp}. 
The relative contributions of the uncorrelated and correlated particle sub-samples have to be determined assuming a realistic two-body interaction providing the two-body correlation. 

In the MC simulations each event is represented by specific values for the momenta $(\mathbf{p}_1,\mathbf{p}_2,\mathbf{p}_3)$ of the three-particles, to be extracted from a \textit{probability distribution function (pdf)} of the momenta. The 
The domain is 
\begin{equation}
\begin{aligned}
    \mathcal{D} = \{ (\mathbf{p}_1, \mathbf{p}_2, \mathbf{p}_3) \in \mathcal{S}\ |\ &|\mathbf{p}_1| < m_1\ \land\ |\mathbf{p}_2| < m_2\ \land\\
    &\land\ |\mathbf{p}_3| < m_3 \} \ ,
\end{aligned}
\end{equation}
chosen to fulfill the non-relativistic condition.
For the uncorrelated particles sub-sample, the components of the momentum vectors are drawn from uniform distributions $U(\mathbf{p}) = U(p_x) U(p_y) U(p_z)$ while the genuine two-particle correlation is obtained using a joint distribution of the corresponding momenta $f_2(\mathbf{p}_{1},\mathbf{p}_{2})$. The global $pdf$ is then defined as
\begin{equation}
\begin{aligned}
    f(\mathbf{p}_1,\mathbf{p}_2,\mathbf{p}_3) =&\ q\ U(\mathbf{p}_1) \ U(\mathbf{p}_2)\ U (\mathbf{p}_3)\ +\\ &+\ p\ f_2 (\mathbf{p}_1, \mathbf{p}_2)\ U (\mathbf{p}_3) \ ,
\end{aligned}
\label{eq:pdf1}
\end{equation}
where $q$ and $p$ are the weights of the two sub-samples and are calculated in Section \ref{sec:Eop}.
The joint $pdf$ which is chosen to generate the two-body correlation is a multivariate Gaussian distribution defined as follows
\begin{equation}
    f_2(\mathbf{p}_{1},\mathbf{p}_{2}) =
 \mathcal{N}  
 \exp\left[{-\frac{\left(\frac{m_2}{m_1 + m_2}\mathbf{p}_{1}- \frac{m_1}{m_1 + m_2}\mathbf{p}_{2}\right)^2}{2\ \sigma_2^2}}\right] 
\ ,
\label{pdfMC}
\end{equation}
where $\mathcal{N}$ is a normalization constant and $\sigma_2$, together with $q$ and $p$, 
represent the parameters of the genuine two-body correlation and have to be determined. 
Since the $pdf$s are normalised to unity in the domain $\mathcal{D}$, the completeness relation $q \ + \ p \ = 1$ is fulfilled.
The choice of the Gaussian distribution in Eq. \eqref{pdfMC} allows to factorise the $pdf$s of the components of the momentum vectors
\begin{equation}
     f_2(\mathbf{p}_{1},\mathbf{p}_{2}) =  f_{2,x}({p}_{1,x},{p}_{2,x})\  f_{2,y}({p}_{1,y},{p}_{2,y})\
     f_{2,z}({p}_{1,z},{p}_{2,z})\ ,
\end{equation}
and to easily generate numerically the vector components which are used to calculate $k^*$ and $Q_3$ event-by-event.

\subsection{Extraction of the parameters}\label{sec:Eop}

Given that particle 3 is uncorrelated, the $pdf$ from Eq. \eqref{eq:pdf1} simplifies to
\begin{equation}
    f(\mathbf{p}_1,\mathbf{p}_2) =\ q\ U(\mathbf{p}_1) \ U(\mathbf{p}_2)\ +\ p\ f_2 (\mathbf{p}_1, \mathbf{p}_2)\ \ .
\label{eq:pdf2}
\end{equation}
Sampling the single particle momenta from $f(\mathbf{p}_1,\mathbf{p}_2)$, the corresponding $k^*$ and $N_\mathrm{same}(k^*)$ can be obtained. 
This sample can be expressed as
\begin{equation}
    N_\mathrm{same}^\mathrm{MC}(k^*) = q \ N_\mathrm{mixed}^\mathrm{MC}(k^*)\ +\ p\ g_2 (k^*) \ ,
    \label{eq:fitfunction}
\end{equation}
where $N_\mathrm{mixed}^\mathrm{MC}(k^*)$ is the uncorrelated, mixed event, distribution and $g_2 (k^*)$ is the function $f_2 (\mathbf{p}_1, \mathbf{p}_2)$ in terms of $k^*$. 

The two-body interaction is commonly modeled using the Lednick{\'y} formalism \cite{Lednicky:1981su}, which can express the correlation function $C_2^\mathrm{Lednick{\text{\'y}}}$ in terms of the scattering length $f_0$, the effective range $d_0$ of the interaction, and the source radius $r_0$. The same event sample can be expressed as 
\begin{equation}
    N_\mathrm{same}(k^*) = C_2^\mathrm{Lednick{\text{\'y}}} (k^*) \ N_\mathrm{mixed}(k^*) \ ,
    \label{eq:Ledni}
\end{equation}
allowing to obtain the parameters 
$q$, $p$ and $\sigma_2$ for a given set of scattering parameter and emission source radius. $N_\mathrm{mixed}(k^*)$ in Eq. \eqref{eq:Ledni} is generated drawing a sample from the $pdf$ in Eq. \eqref{eq:pdf2} with the condition $(q,p)=(1,0)$, corresponding to the hypothesis of no genuine two-body correlation. Finally, the corresponding $N_\mathrm{same}(k^*)$ is evaluated from Eq. \eqref{eq:Ledni}, and the result is fitted with using Eq. \eqref{eq:fitfunction} to extract $q$ and $\sigma_2$ with the constraint $p = 1 - q$.

In order to perform a realistic test, $C_2^\mathrm{Lednick{\text{\'y}}}(k^*)$ is evaluated assuming the scattering parameters corresponding to the spin-0 p$\mathrm{\Lambda}$ interaction calculated within the $\chi$EFT at the next-to-the-leading-order (NLO) given in Ref. \cite{HAIDENBAUER201324} ($f_0 = 2.91$ fm, $d_0 = 2.78$ fm). A source radius $r_0 = 1.25$ fm, typical for small colliding systems at the LHC \cite{2020135849,ALICE:2018ysd}, is assumed. 
\begin{figure}
    \centering
    \includegraphics[width=0.48\textwidth]{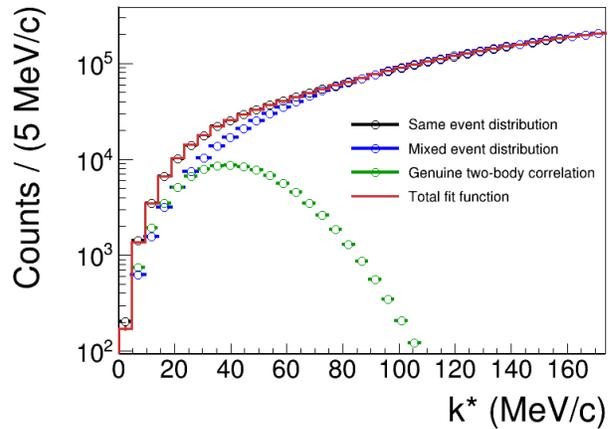}
    \caption{Fit of the same event distribution for the pair (1,2) obtained using the Lednick{\'y} model (black open points). The chosen scattering parameters correspond to the spin-0 p$\mathrm{\Lambda}$ interaction calculated within the $\chi$EFT at the NLO given in Ref. \cite{HAIDENBAUER201324} ($f_0 = 2.91$ fm, $d_0 = 2.78$ fm) and a source radius $r_0 = 1.25$ fm is chosen. The total fit function is the red histogram, the open green and blue points represent the contributions to the fit due to the genuine two-body and mixed event components in the simulated data sample. See text for details.}
    \label{fig:fit}
\end{figure}
In Fig. \ref{fig:fit}, the fit of the $N_\mathrm{same}(k^*)$ distribution (black open points) is shown. The red histogram is the fit function, the open green and blue points represent the contributions to the fit due to the genuine two-body and mixed event components in the simulated data sample. 
The obtained parameters $q$, $p$ and $\sigma_2$ are reported in Table \ref{tab:fit}.

\begin{table}
    \centering
    \caption{Weights $q$, $p$ and $\sigma_2$ parameters extracted from the fit of the same event distribution using Eq. \eqref{eq:fitfunction} (see text for the details).}
    \label{tab:fit}
    \resizebox{\columnwidth}{!}{
    \begin{tabular}{c|c|c}
    $q$ & $p$ & $\sigma_2$ \\
    \hline
        (99.734 $\pm$ 0.002) \% & (0.266 $\pm$ 0.002) \% & ($27.7 \pm 0.1$) MeV/c \\
    \hline
    \end{tabular}}
\end{table}


\subsection{Projected two-body correlation function}

\begin{figure}
    \centering
    \includegraphics[width=0.48\textwidth]{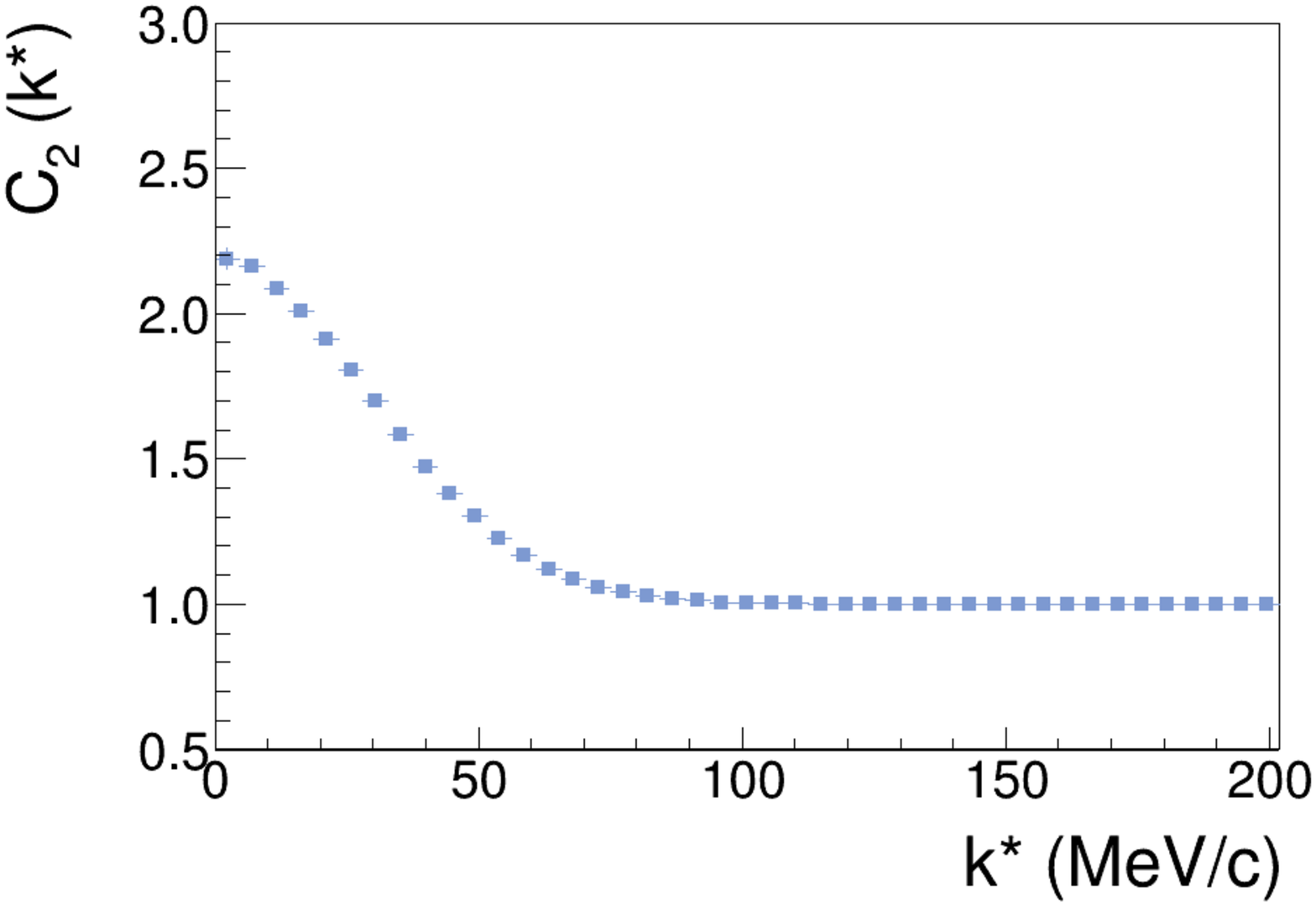}
    \includegraphics[width=0.48\textwidth]{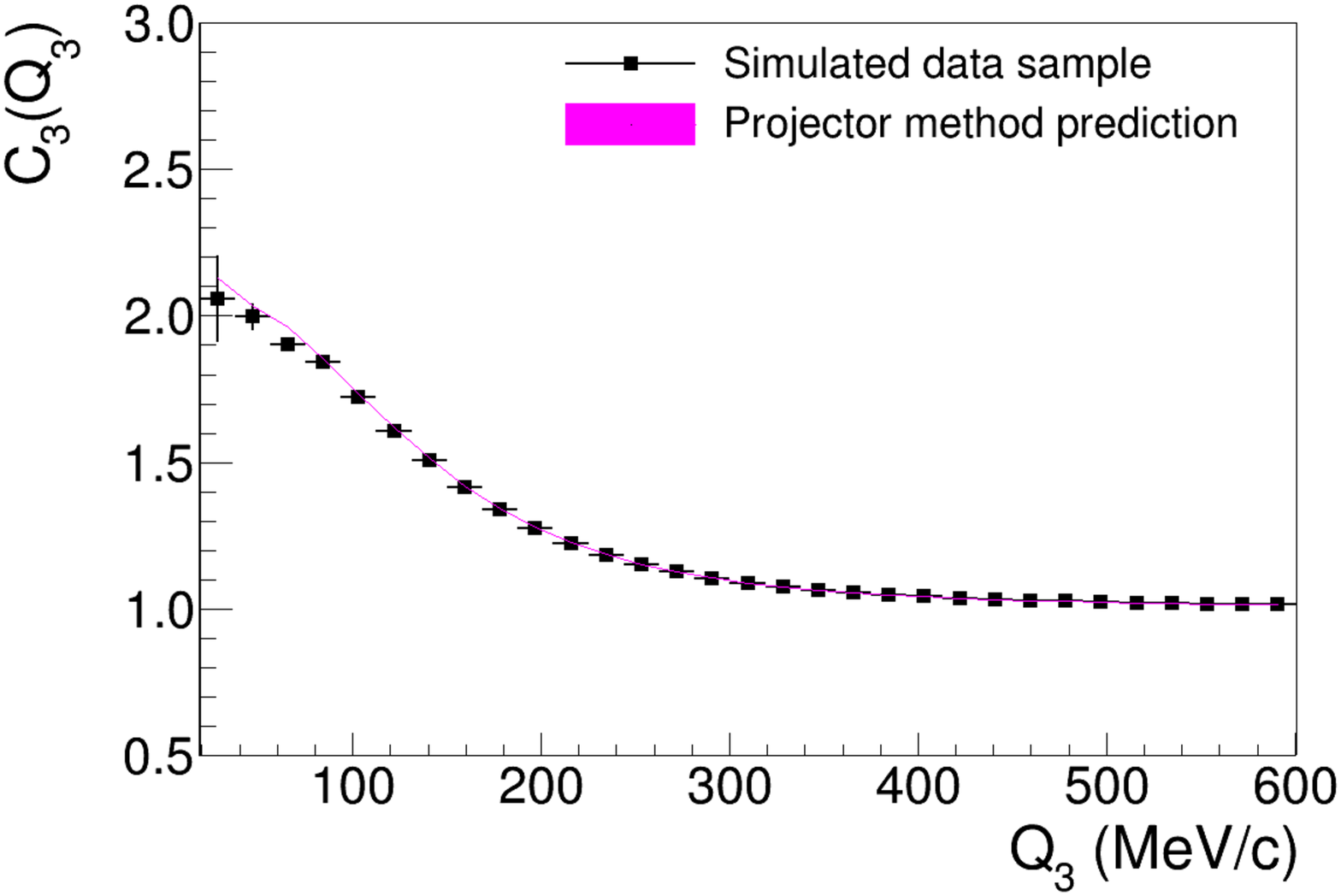}
    \caption{{\em Top panel:} Two-body correlation function for the simulated p$\mathrm{\Lambda}$ data sample as a function of the relative momentum of the interacting pair.\\
    {\em Bottom panel:} Three-body correlation function for the simulated triplets (black boxes) as a function of $Q_3$. Particles (1,2) are correlated (the correlation function is shown in the top panel), particle 3 is not correlated to the (1,2) pair. The uncertainties represent the statistical errors of the MC simulation. The magenta band is obtained by projecting the (1,2) correlation function (top-panel) from $k^{*}$ to the $Q_3$ of the triplet. The width of the band represents the uncertainty obtained from the error propagation of the statistical uncertainties of correlation function in the top panel.}
    \label{fig:MCsimulation}
\end{figure}

Using the weights and the $\sigma_2$ in Table \ref{tab:fit}, the $pdf$ of the particle momenta in Eq. \eqref{eq:pdf1} is defined. The three-body MC data sample, where particle 3 is not correlated to the pair (1,2), is finally generated. 
The correlation functions $C_2(k^{*})$ and $C_3(Q_3)$ are calculated using the experimental formulas in Eqs. \eqref{eq:C2PM} and \eqref{eq:C3}. The obtained correlation functions are shown in Fig. \ref{fig:MCsimulation}, top and bottom panels respectively. 
The uncertainties represent the statistical errors of the MC simulation. The magenta band in the bottom panel is obtained by substituting in Eq. \eqref{kprojQ3} as $C_2(k^{*})$ the two-body correlation shown in the top panel of Fig. \ref{fig:MCsimulation}. The simulated correlation function is in agreement with the correlation function obtained by projecting $k^{*}$ onto $Q_3$, validating the projector method developed in Section \ref{SectionProjector}. The band of the magenta curve represents the statistical uncertainty of the projected correlation function,
which is significantly reduced compared to the simulated data sample. This is related to the fact that two-body correlation $C_2(k^*)$ has smaller uncertainties compared to $C_3(Q_3)$. The projector method allows to evaluate the two-body (background) contributions in the measured three-body correlation with much higher precision, compared to any data-driven determination.

\section{Simulation of a genuine three-body correlation}
\label{Sppp-ppp}
The next step is to consider the complete triplets sample decomposition in Eq. \eqref{eq:KuboDec} in the MC data sample. The total two-body contributions in the simulated three-body correlation function will be modelled using the projector method to calculate $C_{3}^{2b}(Q_3)$ in Eq. \eqref{sumC3}. 
Any deviation from $C_{3}^{2b}(Q_3)$ 
will reflect the non-vanishing three-body cumulant in the simulated triplet sample. 
The study presented in the following aims to provide a first qualitative procedure that could be used to test the available three-body potentials models in view of the future femtoscopic measurements.

\subsection{MC data sample}\label{sec:MCdatasample}

Using the same procedure adopted in Section \ref{sec:datasample} and following the Kubo's decomposition in Eq. \eqref{eq:KuboDec}, the MC triplets data sample is obtained extracting the single particle momenta from the $pdf$
\begin{equation}
\begin{aligned}
    &f(\mathbf{p}_1,\mathbf{p}_2,\mathbf{p}_3) =\ w_U\ U(\mathbf{p}_1) \ U(\mathbf{p}_2)\ U (\mathbf{p}_3)\ +\\ &+\ w_{2B}\ [ f_2 (\mathbf{p}_1, \mathbf{p}_2)\ U (\mathbf{p}_3)\ +\ f_2 (\mathbf{p}_2, \mathbf{p}_3)\ U (\mathbf{p}_1)\ +\\ &+\ \ f_2 (\mathbf{p}_3, \mathbf{p}_1)\ U (\mathbf{p}_2) ] \  +\ w_{3B}\ f_3 (\mathbf{p}_1, \mathbf{p}_2, \mathbf{p}_3) \ ,
\end{aligned}
\label{eq:pdf3B}
\end{equation}
where $w_U$, $w_{2B}$ and $w_{3B}$ are respectively the weights of the uncorrelated, the two-body and three-body correlated sub-samples to be determined; the functions $f_2(\mathbf{p}_i,\mathbf{p}_j)$ are the joint two-particle momenta $pdf$s defined in Eq. \eqref{pdfMC}; $f_3 (\mathbf{p}_1, \mathbf{p}_2, \mathbf{p}_3)$ is the joint $pdf$ of the three-particle momenta, providing the genuine three-body correlation contribution.  

\begin{figure}
    \centering
    \includegraphics[width=0.4\textwidth]{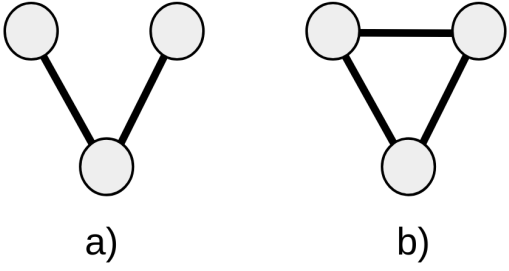}
    \caption{Schematic representation of the three-body interaction model used in the MC simulation. Two cases are considered: a) two pairs in the triplet interact simultaneously; b) all the three pairs interact simultaneously.}
    \label{fig:triplets}
\end{figure}

To determine $f_3 (\mathbf{p}_1, \mathbf{p}_2, \mathbf{p}_3)$ and to fix the weights of the sub-samples in the $pdf$ in Eq. \eqref{eq:pdf3B} a model for the genuine three-body interaction is needed. Since theoretical calculations of three-body correlation functions, involving two- and three-body potentials, are not available in literature, a toy model will be considered in the following. A genuine three-body correlation 
is obtained if, in the same event, at least two pairs of the triplet are correlated. 
Each pair correlation is assumed to be given by the same parameters of the genuine two-body interaction. Consequently, in the adopted toy model the three-body interaction occurs via a simultaneous two-body interaction. Since there are no spectator particles in the triplet, such correlation gives a higher order contribution in the Kubo's decomposition rule, namely the genuine three-body term of the cumulant expansion. A similar treatment of the three-body forces is given in Ref. \cite{Lonardoni:2014bwa}, where the three-body potential is formalised in terms of multiple Yukawa-like potentials. 
Such condition is satisfied in two cases: a) only two pairs in the triplet interact; b) all the three pairs in the triplet interact. A schematic representation is shown in Fig. \ref{fig:triplets}.
Under this hypothesis it is possible to define consistently the weights and the function $f_3 (\mathbf{p}_1, \mathbf{p}_2, \mathbf{p}_3)$, given a two-body interaction model. 

\subsection{Extraction of the weights}\label{sec:w3B}

The weights $w_U$, $w_{2B}$, $w_{2B}$ and the function $f_3$ are obtained using the same three particles considered in Section \ref{SMC}, with mass of about 1 GeV and two-body interaction parameters $q$, $p$ and $\sigma_2$ obtained from the fit in Section \ref{sec:Eop} and given in Table \ref{tab:fit}. Each of the three pairs in the triplet, namely the pairs (1,2), (2,3) and (3,1), is correlated or not correlated in the momentum space. The probability to find a correlated $(i,j)$ pair in the generated data sample is given by the parameter $p$ and, consequently, the probability to find an uncorrelated $(i,j)$ pair in the same data sample is given by $q$. The weight $w_U$ of the uncorrelated triplets in the MC sample is then calculated as the probability that the three pairs are not correlated, i.e.
\begin{equation}
    w_U = q \cdot q \cdot q = (99.20 \pm 0.06) \% \ .
\end{equation}
The weight of the two-body correlated samples is given by the probability that only one pair interacts
\begin{equation}
    w_{2B} = p \cdot q \cdot q = (0.265 \pm 0.002 ) \% \ .
\end{equation}
Including the three possible permutations the global two-body contribution amounts to $(0.794 \pm 0.007) \%$.
For the three-body interaction, the separated weights for the cases a) and b) are evaluated. The weight for the case a) is
\begin{equation}
    w_{3B}^{a)} = p \cdot p \cdot q = (7.1 \pm 0.1) \times 10^{-6}
\end{equation}
and for the case b) 
\begin{equation}
    w_{3B}^{b)} = p \ p \ p = (1.88 \pm 0.04) \times 10^{-8} \ .
\end{equation}
The global genuine three-body correlations amounts to
\begin{equation}
    w_{3B} = 3 \ w_{3B}^{a)} + w_{3B}^{b)} = (2.12 \pm 0.03) \cdot 10^{-5}
\end{equation}
where the factor 3 accounts for the permutations of the pair combinations in the case a).

To define the $pdf$ of the genuine three-body correlation $f_3 (\mathbf{p}_1, \mathbf{p}_2, \mathbf{p}_3)$ a correspondence criterion between the weights and the distributions is applied. Consistently with the definitions given in Section \ref{SMC}, the $q$ parameter corresponds to uniform distributions and the $p$ parameter to the multivariate Gaussian $f_2 (\mathbf{p}_1, \mathbf{p}_2)$: 
\begin{equation}
    \begin{aligned}
        &q \qquad \longrightarrow \qquad U(\mathbf{p}_1)\ U(\mathbf{p}_2) \\
        &p \qquad \longrightarrow \qquad f_2(\mathbf{p}_1, \mathbf{p}_2) \ .
    \end{aligned}
\end{equation}
Accordingly, for the three-body correlation in the case a) the weight $w_{3B}^{a)} = p \cdot p \cdot q$ corresponds to the $pdf$
    \begin{equation}
    \begin{aligned}
        f_3^{a)}(\mathbf{p}_1,\mathbf{p}_2,\mathbf{p}_3) = f_2&(\mathbf{p}_1, \mathbf{p}_2) \ f_2(\mathbf{p}_2, \mathbf{p}_3)\ U(\mathbf{p}_3)\ U(\mathbf{p}_1) + \\
        + U&(\mathbf{p}_1)\ U(\mathbf{p}_2)\ f_2(\mathbf{p}_2, \mathbf{p}_3) \ f_2(\mathbf{p}_3, \mathbf{p}_1) + \\
        + f_2&(\mathbf{p}_1, \mathbf{p}_2)\ U(\mathbf{p}_2)\ U(\mathbf{p}_3) \ f_2(\mathbf{p}_3, \mathbf{p}_1) = \\
        = \mathcal{N}& \exp\left[{-\frac{\left(\frac{m_2}{m_1 + m_2}\mathbf{p}_{1}- \frac{m_1}{m_1 + m_2}\mathbf{p}_{2}\right)^2}{2\ \sigma_2^2}}\right]\\ 
        & \exp\left[{-\frac{\left(\frac{m_3}{m_2 + m_3}\mathbf{p}_{2}- \frac{m_2}{m_2 + m_3}\mathbf{p}_{3}\right)^2}{2\ \sigma_2^2}}\right] + \\
        +\ &\mathrm{permutations} \ ,
    \end{aligned}
        \end{equation}
where $\mathcal{N}$ is a normalisation constant and $\sigma_2$ is given in Table \ref{tab:fit}. For the three-body correlation in the case b) the weight $w_{3B(b)} = p \ p \ p$ corresponds to the $pdf$
    \begin{equation}
    \begin{aligned}
        f_3^{b)}(\mathbf{p}_1,\mathbf{p}_2,\mathbf{p}_3) = f_2&(\mathbf{p}_1, \mathbf{p}_2) \ f_2(\mathbf{p}_2, \mathbf{p}_3)\ f_2(\mathbf{p}_3, \mathbf{p}_1) = \\ 
        = \mathcal{N}& \exp\left[{-\frac{\left(\frac{m_2}{m_1 + m_2}\mathbf{p}_{1}- \frac{m_1}{m_1 + m_2}\mathbf{p}_{2}\right)^2}{2\ \sigma_2^2}}\right] \\ & \exp\left[{-\frac{\left(\frac{m_3}{m_2 + m_3}\mathbf{p}_{2}- \frac{m_2}{m_2 + m_3}\mathbf{p}_{3}\right)^2}{2\ \sigma_2^2}}\right] \\ & \exp\left[{-\frac{\left(\frac{m_1}{m_1 + m_3}\mathbf{p}_{3}- \frac{m_3}{m_1 + m_3}\mathbf{p}_{1}\right)^2}{2\ \sigma_2^2}}\right] \ .
    \end{aligned}
    \label{eq:pdf3b}
    \end{equation}
In the MC triplets sample, the contribution due to the genuine three-body correlation is finally drawn from the following weighted distribution
\begin{equation}
\begin{aligned}
     f_3 (\mathbf{p}_1, \mathbf{p}_2, \mathbf{p}_3) &= \frac{w_{3B}^{a)}}{w_{3B}}\ f_3^{a)} (\mathbf{p}_1, \mathbf{p}_2, \mathbf{p}_3)\ +\\
     &+\ \frac{w_{3B}^{b)}}{w_{3B}}\ f_3^{b)} (\mathbf{p}_1, \mathbf{p}_2, \mathbf{p}_3) \ .
\end{aligned}
\end{equation}



\subsection{Results}

\begin{figure}
    \centering
    \includegraphics[width=0.48\textwidth]{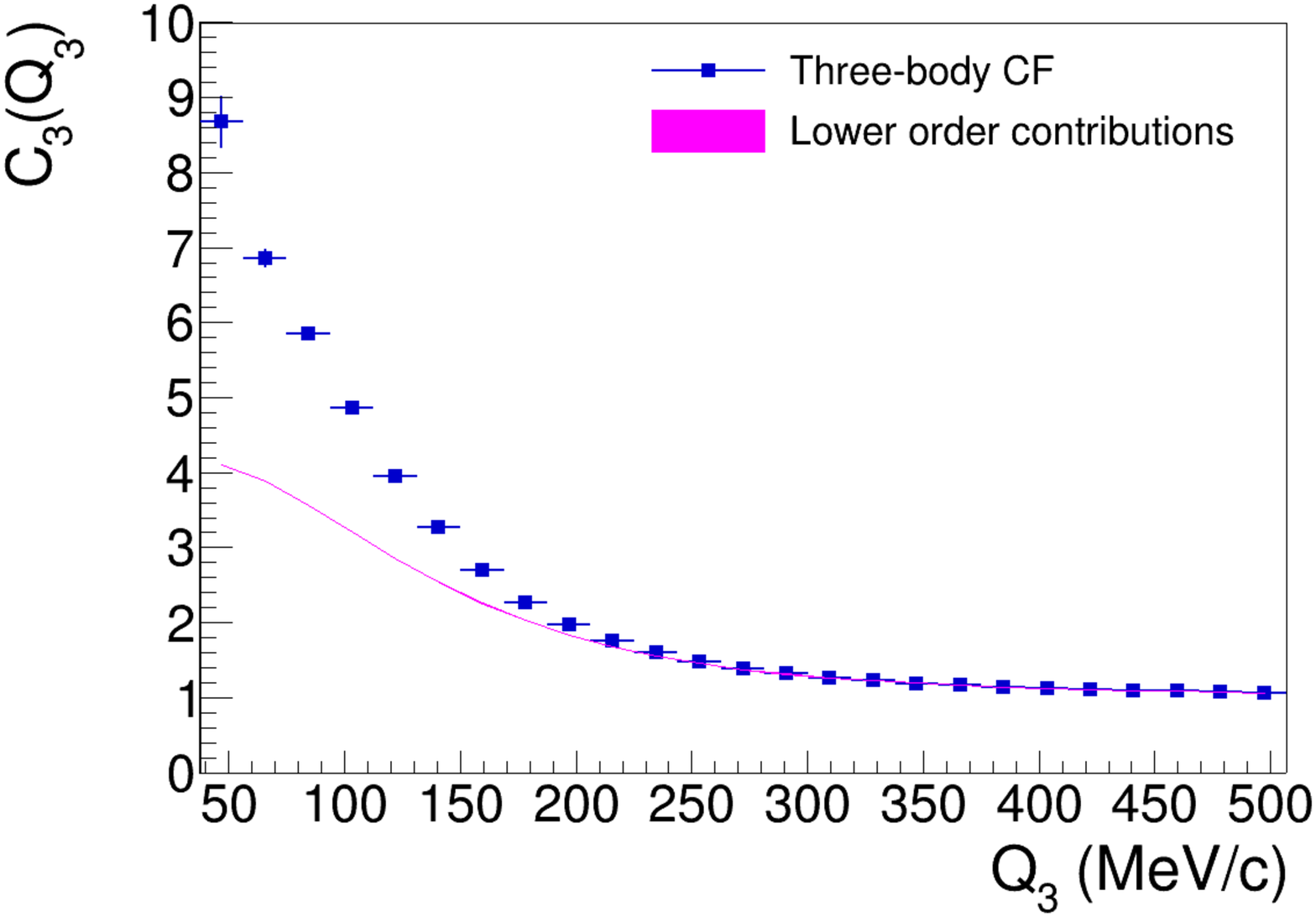}
    \includegraphics[width=0.48\textwidth]{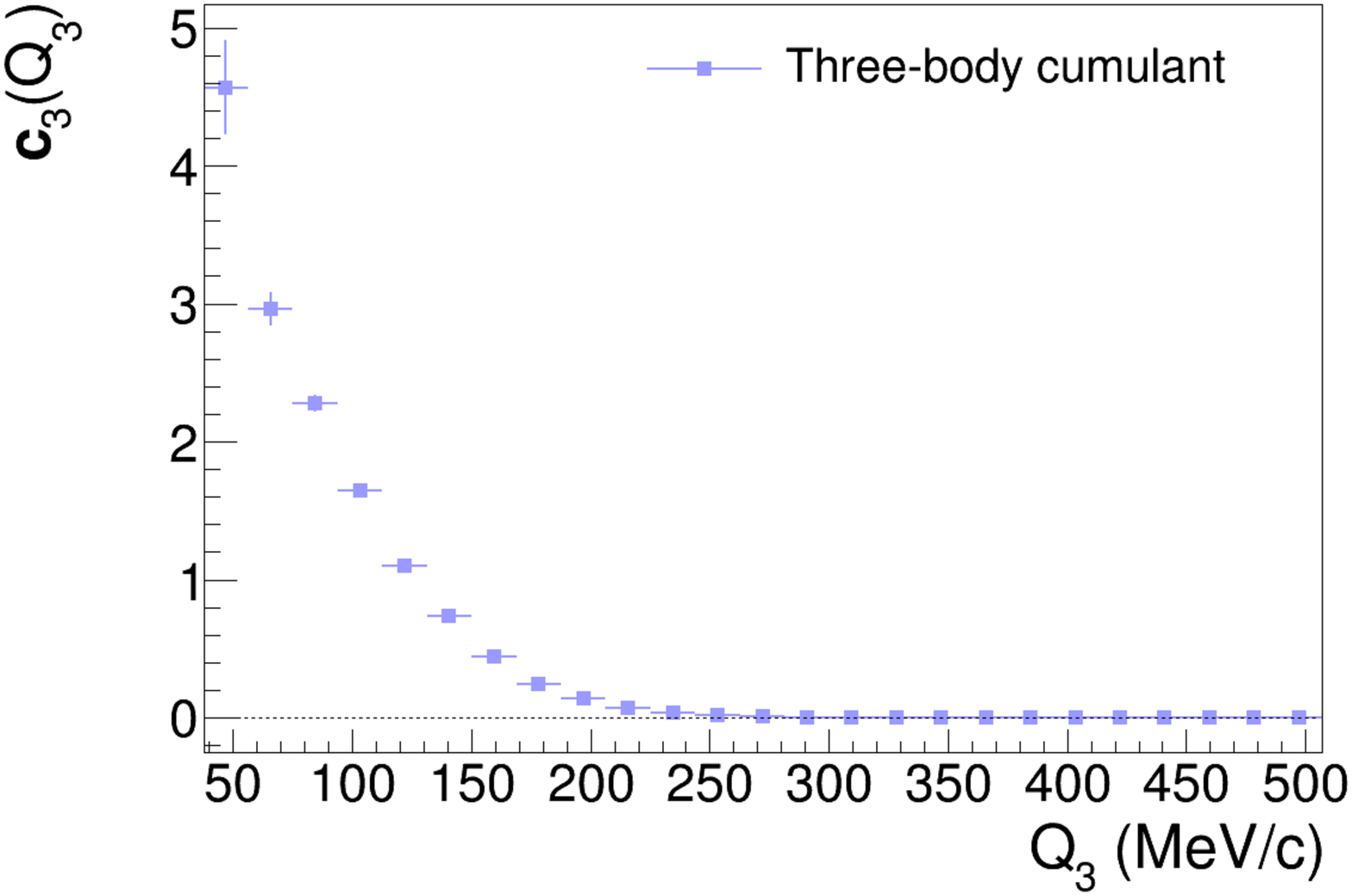}
    \caption{{\it Top panel:} Three-body correlation function for the MC simulated triplets (blue boxes) as a function of the hyper-momentum $Q_3$. Genuine two-body and three-body correlations are considered in simulated data sample. The error bars represent the statistical uncertainties of the MC simulation. The total two-body interaction contribution calculated using the projector method is represented by the magenta band, the width is the corresponding statistical uncertainty. \\
    {\it Bottom panel:} Three-body cumulant for the MC simulated triplets obtained by subtracting the three-body correlation function and the lower order contributions calculated with the projector method. 
    }
    \label{fig:3B27R}
\end{figure}

The three-body correlation function obtained from the MC simulation, where the genuine three-body correlation is included, is shown in Fig. \ref{fig:3B27R} top panel with blue boxes. 
The magenta band represents the total lower order contributions $C_{3}^{2b}(Q_3)$ in the correlation function and it is obtained applying Eq. \eqref{sumC3} the two-body correlation function shown in Fig. \ref{fig:MCsimulation} (top panel) projected onto $Q_{3}$. 
The width of the magenta band is the corresponding statistical uncertainty. A deviation of the simulated correlation function with respect to the two-body interaction contribution appears in the low $Q_3$ region. 
The femtoscopic cumulant is obtained from Eq. \eqref{cumulantsCF} as a function of $Q_3$
\begin{equation}
    \mathbf{c}_{3}(Q_{3}) = C_{3}(Q_{3}) - C_{3}^{2b}(Q_3) \ ,
\end{equation}
where $C_{3}(Q_{3})$ is the simulated three-body correlation function and it is shown in Fig. \ref{fig:3B27R} bottom panel. The three-body cumulant is effectively zero for $Q_3 > 300$ MeV/c. 
In the adopted toy model for the genuine three-body correlations, the simultaneous interaction of the pairs in the triplets occurs with the same strength as in the case of a genuine two-body interaction. 

\begin{figure}
    \centering
    \includegraphics[width=0.48\textwidth]{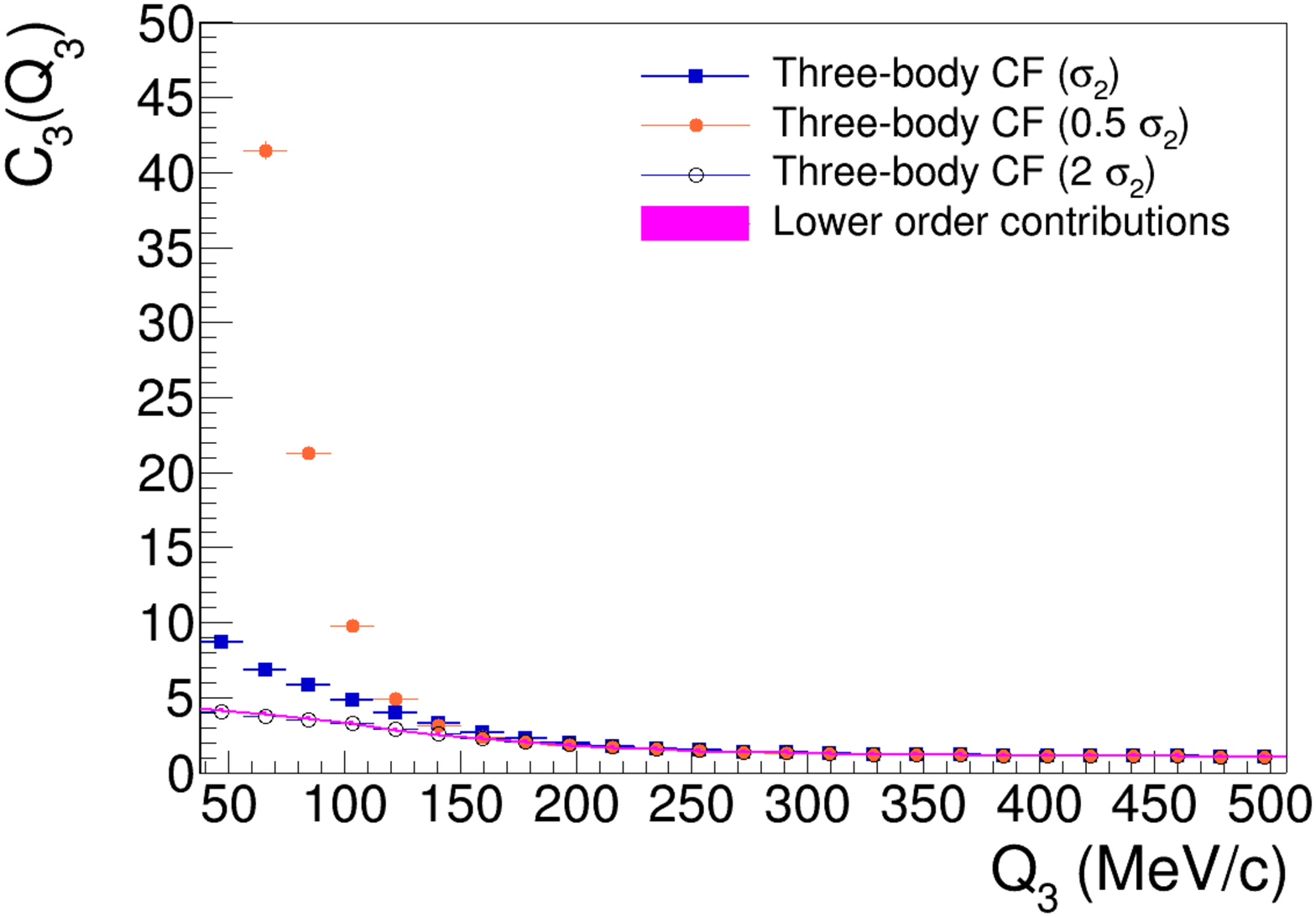}
    \includegraphics[width=0.48\textwidth]{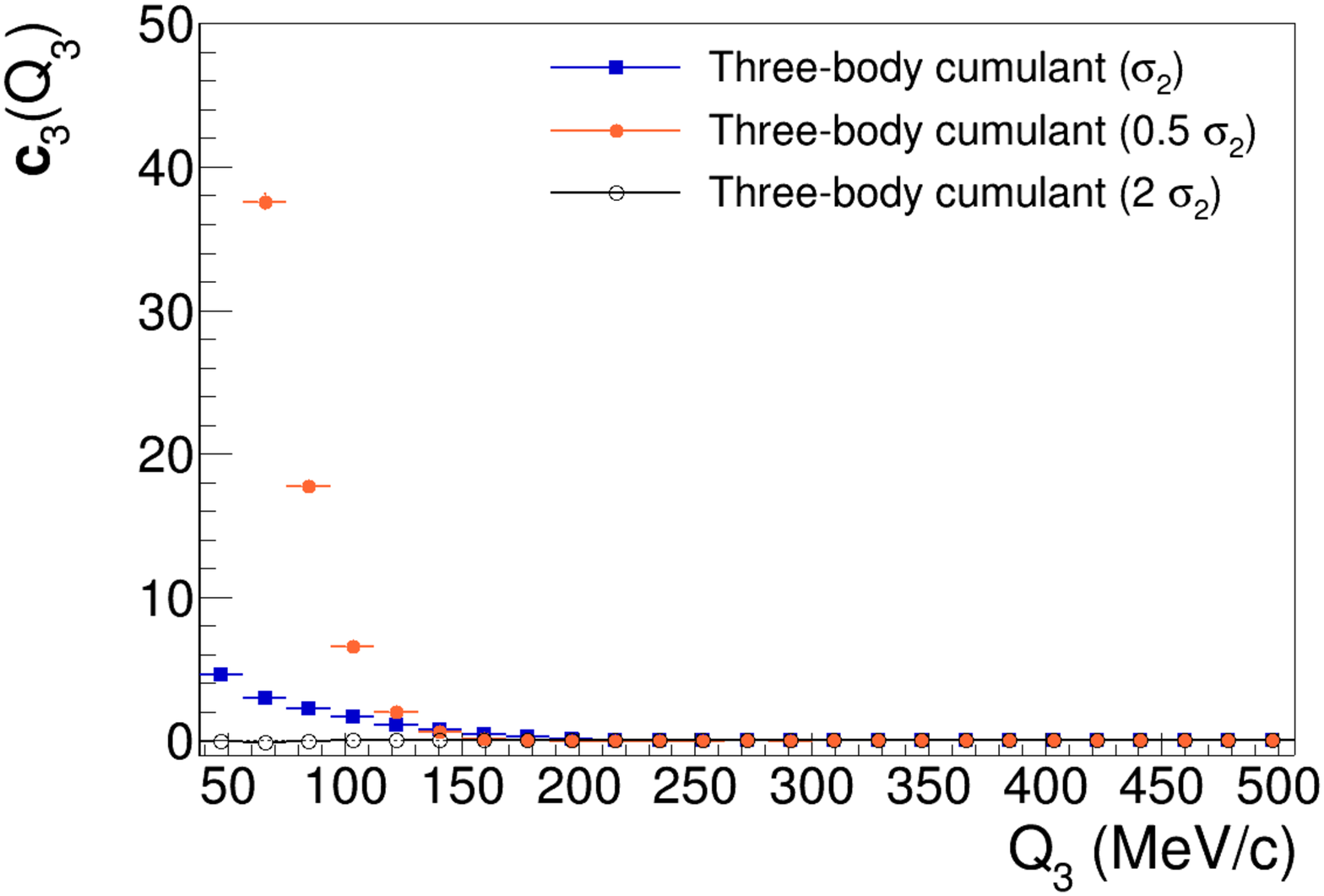}    
    \caption{{\it Top panel:} Three-body correlation functions obtained from MC simulations of triplet samples where the genuine two- and three-particle correlations are included. 
    Blue boxes correspond to the case where the simultaneous pairwise interactions used to simulate the three-body forces have the same strength of the genuine two-body interaction ($\sigma = \sigma_2$), full orange points correspond to the case of stronger interaction ($\sigma = 0.5  \sigma_2$) and open black points to the case weaker interaction ($\sigma = 2  \sigma_2$). See text for details. The magenta band represents the contributions to the correlation function due to the lower order contributions obtained using the projector method.
    {\it Bottom panel:} Three-body cumulant for the MC simulated triplets obtained by subtracting the three-body correlation function and the lower order contributions calculated with the projector method. The same color code used in the plot in the top panel is used to distinguish the cases of $\sigma = 0.5  \sigma_2 \ ,\ \sigma_2\ ,\ 2  \sigma_2$.
    }
    \label{fig:MC12}
\end{figure}
In the following, the strength of the three-body correlation is varied. To this end, the $\sigma$ parameter in the Gaussian functions in $f_3(\mathbf{p}_1,\mathbf{p}_2,\mathbf{p}_3)$ is considered to be not identical to the $\sigma_2$ parameter of the two-body interaction ($\sigma \neq \sigma_2$). The cases of $\sigma = 0.5\ \sigma_2$ and $\sigma = 2\ \sigma_2$ are explored, corresponding respectively to an increased or decreased strength of the interaction.
The resulting correlation functions and three-body cumulants are shown in Fig. \ref{fig:MC12}. 
As expected, decreasing the strength of the simultaneous pairwise correlations (using $\sigma = 2 \sigma_2$) the shape of the correlation function (open black points in Fig. \ref{fig:MC12}) approaches the shape of the predicted lower order contributions. On contrary, increasing the strength of the simultaneous pairwise correlations (using $\sigma = 0.5 \sigma_2$) the shape of the correlation function deviates more strongly from the lower order contributions in the low $Q_3$ region. 

Although the study conducted in this paper is qualitative, it is shown that by measuring the three-hadron correlation function and using the Kubo's cumulant method it is possible to infer on the parameters of the interaction models. 
It has been demonstrated that if the interaction strength and range of the three- and two-body forces are similar, the genuine three body signal observed in the cumulant becomes significant, and likely detectable by the LHC experiments planned in the near future. 

\section{pK$^+$K$^-$ combinatorial background}\label{Spphi}
\begin{figure}[!h]
    \centering
    \includegraphics[width=0.48\textwidth]{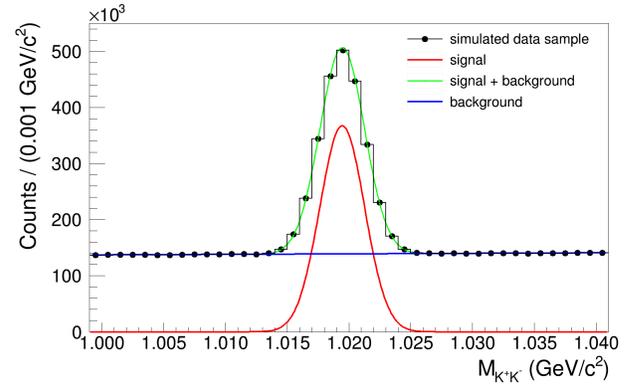}
    \caption{Simulated K$^+$K$^-$ invariant mass spectrum (black points and histogram). The signal (red distribution) is simulated using a Gassian function with center and full width at half maximum equal to the mass and the decay width of the $\upphi$ meson from the PDG. A first order polynomial is assumed for the combinatorial background (blue distribution). The sum of the signal and background contribution is also shown in green.}
    \label{fig:minvkpkm}
\end{figure}

The method described in Section \ref{SectionProjector} can be used to predict the contribution to the hadron-hadron correlation functions which is due to the combinatorial background affecting the particle identification of the candidate hadrons \cite{acharya2020investigation}. Particles that are reconstructed from the decay channels, such as neutral hadrons, are identified by applying a selection cut in the invariant mass distributions of the decay products. In the selected invariant mass window particles which are not produced in the decay are also selected, entering as a background in the measured correlation function. The case of the $\upphi$ meson, reconstructed through the $\upphi \rightarrow \mathrm{K^+K^-}$ decay, is now considered. An example of the reconstructed M$_\mathrm{K^+K^-}$ invariant mass distribution is shown in Fig. \ref{fig:minvkpkm} for a simulated data sample. The red distribution is the simulated signal for the $\upphi$ and the blue distribution represents the combinatorial background, i.e. the K$^+$K$^-$ pairs which are not emitted in the decay of the $\upphi$.      

Let us suppose now to be interested in the measurement of the genuine p$\upphi$ correlation function and that the $\upphi$ candidates are selected applying the cut $m_{\upphi} - \delta < \mathrm{M}_\mathrm{K^+K^-} < m_{\upphi} + \delta$ to the spectrum in Fig. \ref{fig:minvkpkm}, where $m_{\upphi}$ is the nominal mass of the $\upphi$ and $\delta$ is a chosen window interval. The measured p$\upphi$ correlation function will be affected by the contribution due to the pK$^-$, pK$^+$ and K$^+$K$^-$ interactions. These contributions are commonly evaluated by selecting proper side-bands windows in the invariant mass spectrum in Fig. \ref{fig:minvkpkm}, namely selecting the kaons from regions which are outside the window used for the $\upphi$ particle identification. For the selected pK$^+$K$^-$ triplets the correlation function is evaluated as a function of the relative p$\upphi$ momentum defined as
\begin{equation}
    \mathbf{k}_\mathrm{p\upphi} = \frac{m_{\upphi}}{m_\mathrm{p} + m_\upphi} \mathbf{p}_\mathrm{p} - \frac{m_\mathrm{p}}{m_\mathrm{p} + m_\upphi} (\mathbf{p}_\mathrm{K^+} + \mathbf{p}_\mathrm{K^-}) \ .
\end{equation}

\begin{figure}
    \centering
    \includegraphics[width=0.48\textwidth]{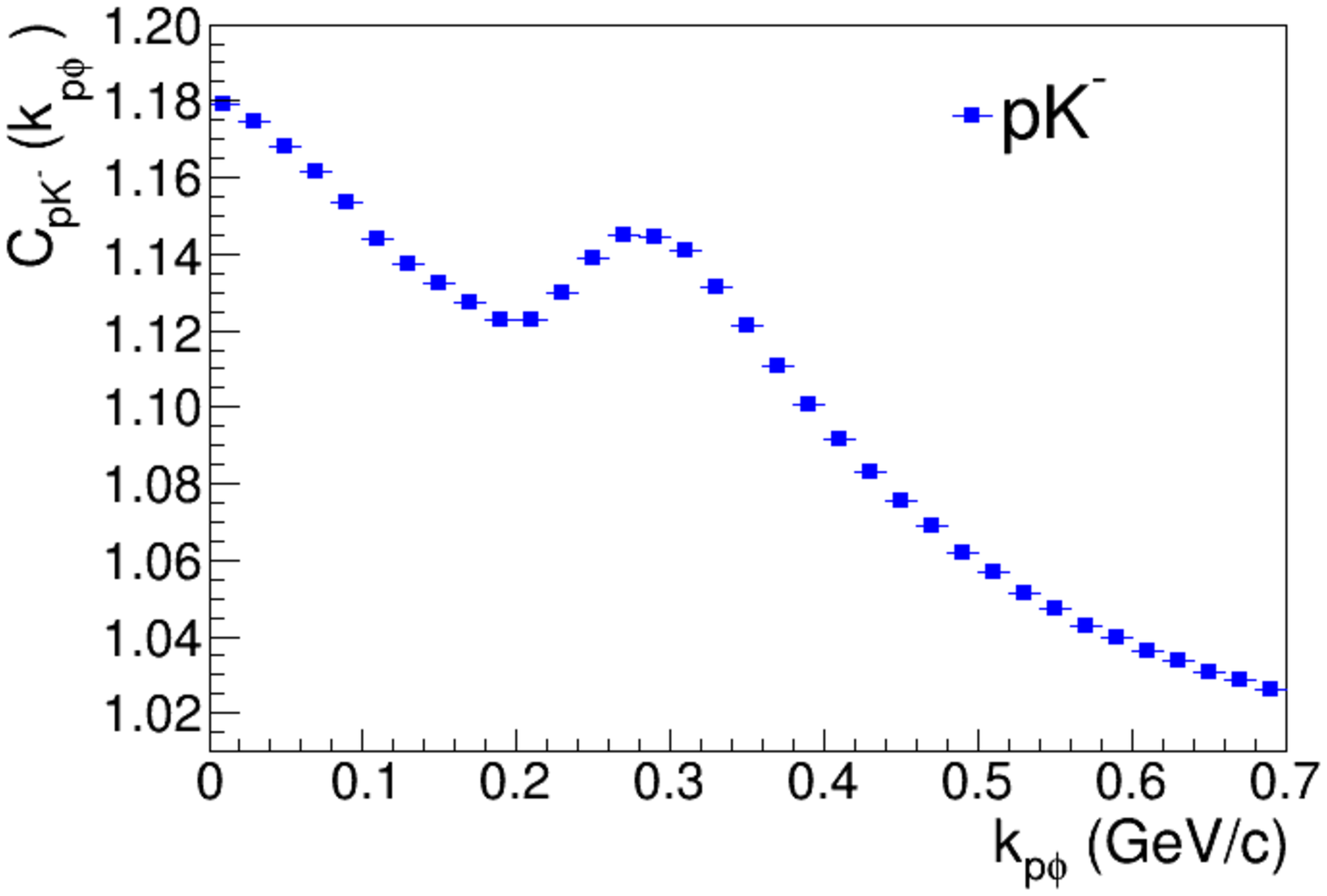}
    \includegraphics[width=0.48\textwidth]{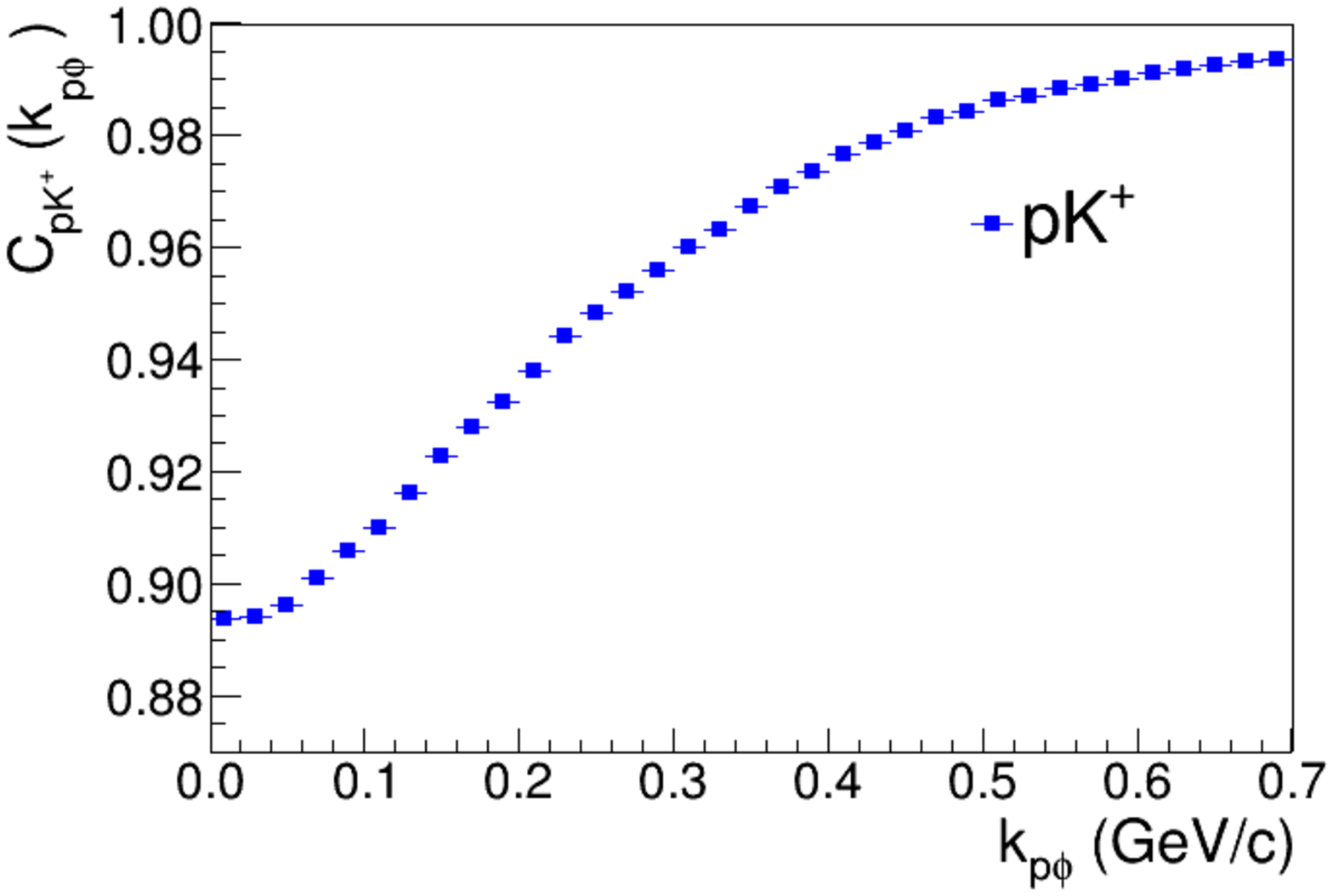}
    \caption{Two-body pK$^-$ (top panel) and pK$^+$ (bottom panel) correlation functions projected onto the $k_\mathrm{p\upphi}$.
    For $C_\mathrm{pK^-}(k^{*})$ and $C_\mathrm{pK^+}(k^{*})$ as functions of $k^{*}$, the experimental distributions measured by ALICE in Ref. \cite{Acharya:2019bsa} are considered. The details of the projections onto $k_\mathrm{p\upphi}$ are given in the text.}
    \label{fig:pKp-pKm}
\end{figure}

\begin{figure}
    \centering
    \includegraphics[width=0.48\textwidth]{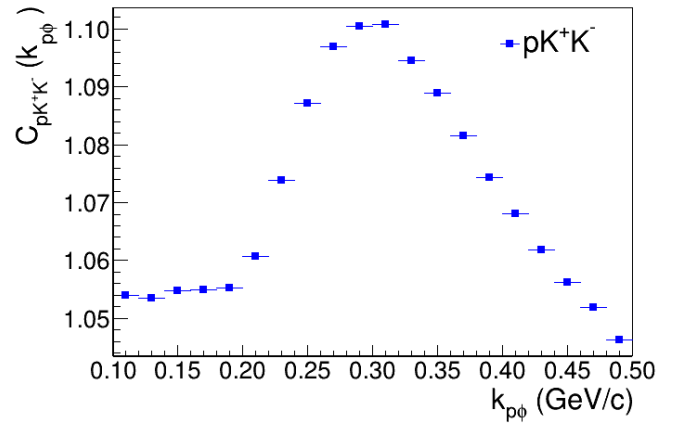}
    \caption{Contribution to the p$\upphi$ correlation function due to the combinatorial pK$^+$K$^-$ background.
    The prediction is calculated using the decomposition $C_\mathrm{pK^+K^-}(k_\mathrm{p\upphi}) = C_\mathrm{pK^-} (k_\mathrm{p\upphi}) + C_\mathrm{pK^+} (k_\mathrm{p\upphi}) - 2$, where $C_\mathrm{pK^-} (k_\mathrm{p\upphi})$ and $C_\mathrm{pK^+} (k_\mathrm{p\upphi})$ are shown in Fig. \ref{fig:pKp-pKm}.}
    \label{fig:combinatorial}
\end{figure}

The same contribution can be calculated by using the formalism developed in Section \ref{SectionProjector}. Given the two-body pK$^-$, pK$^+$ and K$^+$K$^-$ correlation functions as a function of the relative momentum of the interacting pairs, the projection onto the p$\upphi$ relative momentum can be calculated using Eq. \eqref{sumC3}, by replacing $Q_3$ with $k_\mathrm{p\upphi}$, i.e.
\begin{equation}
\begin{aligned}
    C_\mathrm{pK^+K^-}(k_\mathrm{p\upphi}) &= C_\mathrm{pK^-} (k_\mathrm{p\upphi}) + C_\mathrm{pK^+} (k_\mathrm{p\upphi}) + \\
    &+ C_\mathrm{K^+K^-} (k_\mathrm{p\upphi}) - 2 \ ,
\end{aligned}
    \label{Cpphi}
\end{equation}
where the contribution due to the genuine three-body pK$^+$K$^-$ correlation is assumed to be negligible. 
The triplet $\mathrm{(p,K^+,K^-)}$ will be denoted in the following with the numbers $(1,2,3)$ for simplicity. Each contribution in the sum in Eq. \eqref{Cpphi} is evaluated as follows
\begin{equation}
\resizebox{0.9\hsize}{!}{$
    C_{ij}(k_\mathrm{p\upphi}) = \iiint_{(\mathbf{p}_1, \mathbf{p}_2, \mathbf{p}_3) \in \mathcal{D}} C_{ij}(\mathbf{p}_i, \mathbf{p}_j)\ \mathcal{N} \ d^3\mathbf{p}_1 \ d^3\mathbf{p}_2\  d^3\mathbf{p}_3 \ ,
    $}
    \label{intCpphi}
\end{equation}
where the integration domain $\mathcal{D}$ is defined by the kinematic constraints, i.e. $k_{p\upphi} = \mathrm{constant}$ and $m_{\upphi} - \delta < \mathrm{M_{K^+K^-}} < m_{\upphi} + \delta$. The genuine two-body pK$^+$ and pK$^-$ correlation function measured by the ALICE collaboration \cite{Acharya:2019bsa} are used. Since the invariant mass distribution for the combinatorial background is not uniform, each value of $\mathrm{M_{K^+K^-}}$ is weighted with the corresponding shape normalised to the unity in the $\upphi$ candidate window. The shape is given by the background component in the fit of the invariant mass distribution (blue curve in Fig. \ref{fig:minvkpkm}), and the weight turn to be
\begin{equation}
    f_{bkg} (\mathrm{M_{K^+K^-}}) = \frac{\mathrm{Fit}_{bkg}(\mathrm{M_{K^+K^-}})}{\int_{m_{\upphi}- \delta}^{m_{\upphi} + \delta} \mathrm{Fit}_{bkg}(\mathrm{M_{K^+K^-}}) \ d\mathrm{M_{K^+K^-}}} \ .
\end{equation}
The integral in Eq. \eqref{intCpphi} is then performed in two-steps
\begin{enumerate}
    \item a first integration is done in the domain
    \begin{equation}
    \begin{aligned}
         \mathcal{D} = \{ (\mathbf{p}_1, \mathbf{p}_2, \mathbf{p}_3) \in \mathcal{S}\ |\ &k_\mathrm{p\upphi} = \mathrm{const.} \ \land \\
         &\mathrm{M_{K^+K^-}} = \mathrm{const.} \} \ ,
    \end{aligned}
         \label{domain2}
    \end{equation}
    which provides a projected correlation function at a fixed value of the K$^+$K$^-$ invariant mass $C_{ij}(k_\mathrm{p\upphi}; \mathrm{M_{K^+K^-}})$;
    \item each value of the obtained correlation function is weighted by the function $f_{bkg} (\mathrm{M_{K^+K^-}})$
    \begin{equation}
        \int_{m_{\upphi}- \delta}^{m_{\upphi} + \delta} C_{ij}(k_\mathrm{p\upphi}; \mathrm{M_{K^+K^-}}) \ f_{bkg} (\mathrm{M_{K^+K^-}}) \ d\mathrm{M_{K^+K^-}} \ .
    \end{equation}
\end{enumerate}
The pK$^+$ and pK$^-$ correlation functions projected onto $k_\mathrm{p\upphi}$ are shown in Fig. \ref{fig:pKp-pKm} top and bottom respectively. A selection window $\delta = 8\ \mathrm{MeV/c^2}$ is chosen.
The K$^+$K$^-$ correlation function is flat when projected onto $k_\mathrm{p\upphi}$. The reason is that the conditions in Eq.  
\eqref{domain2} are satisfied only for the value of the K$^+$K$^-$ relative momentum given by
\begin{equation}
    {k}_{23} = \frac{1}{2} \ \sqrt{\mathrm{M^2_{K^+K^-}} - (m_2 + m_3)^2} \ ,
\end{equation}
and consequently the projected correlation function is constant.
The K$^+$K$^-$ two-body interaction does not affect the pK$^+$K$^-$ correlation function in Eq. \eqref{Cpphi}. This result is intuitive. Since we are projecting all the two-body interaction contributions onto the relative momentum of the proton with respect to the K$^+$K$^-$ CM momentum, and since the interaction among the kaons does not affect the pair CM kinematics, nor the proton dynamics, the correlation function as a function of $k_\mathrm{p\upphi}$ is flat.
Finally, the prediction for the combinatorial background correlation function in the $\upphi$ region is shown in Fig. \ref{fig:combinatorial}.
The correlation function in Fig. \ref{fig:combinatorial} is obtained with negligible uncertainties. The error in each bin is obtained from the error propagation of the statistical errors for the correlation functions measured in Ref. \cite{Acharya:2019bsa}. This is the main advantage of the method with respect to the side-bands technique. The only source of uncertainties will be represented by the statistical error for the parameters of the fit which are used to set the shape $f_{bkg}$ and that are not considered in this work. 

As a final remark, the shapes of the obtained correlation functions shown in Figs. \ref{fig:pKp-pKm} and \ref{fig:combinatorial} depend on the combinatorial background distribution in the K$^+$K$^-$ invariant mass spectrum. In this case a first order polynomial was chosen and a selection window of 8 MeV/c$^2$ was considered. However the discussed procedure can be applied to any specific analysis under study.

\section{Conclusions}\label{s:conclusion}
In this paper, a formalism for the evaluation of lower order contribution in many-body correlation functions was developed and applied to the three-body case. The method relies on kinematic transformations on the momentum phase space and on the Kubo's cumulants expansion formalism. The kinematic transformations allow to project correlation functions onto the kinematic variable of interest for the considered system and the Kubo's decomposition provides a summing rule for the statistical sub-samples terms that contribute to the total many-particle sample. Considering the decomposition in the case of a three-body statistical sample, the lower order correlations are provided only by pairwise interactions, the remaining particle is not correlated to the interacting pair. Given this assumption it was possible to calculate a projector for the two-body correlation functions onto the kinematic variable $Q_3$ of the three-body system. The method was validated performing a toy Monte Carlo simulation of a three-body system where only two particles were correlated in the momentum space. As a further step, the genuine three-body correlations were included in the Monte Carlo data sample employing a three-body interaction model based on simultaneous pairwise interactions for the pairs in the triplets. The p$\mathrm{\Lambda}$ (S=0) scattering parameters with $\chi$EFT NLO calculations are used to constrain the model. 
Comparing the shape of the generated three-body correlation function with the lower order, two-body, contributions a deviation due to the non-vanishing three-body cumulant appears in the region $Q_3 < 300$ MeV/c. 
The developed method has further applications, it was indeed shown that the same procedure can be applied to predict the pK$^+$, pK$^-$ and K$^+$K$^-$ residual contributions in the p$\upphi$ correlation functions, providing an alternative procedure to the methods  commonly used in the data analyses to remove such source of background.

The main advantage of this method is that the required input are the known, or measured with high precision, two-body correlation functions. This leads to a reduced uncertainty of the projection as compared to the traditional experimental methods used for a direct determination. This will greatly aid the precision of the planned accelerator based experiments measuring the genuine three-body interaction. 

\begin{acknowledgements}
The authors gratefully acknowledge Ante Bilandzic, Emma Chizzali, Andrea Dainese, Norbert Kaiser and Andreas Mathis for the clarifying discussions that helped the improvement of the paper.
\end{acknowledgements}
\appendix
\section{$N$-body Jacobi coordinates}\label{formalisation}
Let us consider a system of $N$ non-relativistic particles of masses $m_i$ and positions $\mathbf{x}_i$ (with $i=1,...,N$) in the 3-dimensional space. 
If no interaction among the particles is assumed, the Hamiltonian is given by the sum of the kinetic terms 
\begin{equation}
    \mathcal{H} = \sum_{i=1}^N \mathcal{T}_i = \sum_{i=1}^N \frac{\mathbf{p}_i^2}{2 \ m_i} \ ,
    \label{Hnbody}
\end{equation}
where $\mathbf{p}_i$ are the kinetic momenta of the $N$ particles. We choose a set of Jacobi coordinates $\mathbf{r}_i$  
defined as follows
\begin{equation}
    \mathbf{r}_i = \mathbf{x}_{i+1} - \frac{\sum_{j=1}^i m_j \ \mathbf{x}_{j}}{\sum_{j=1}^i m_j} \ .
\end{equation}
The corresponding kinetic momenta  $\mathbf{k}_i$ are 
\begin{equation}
\begin{aligned}
        \mathbf{k}_i &= \frac{\sum_{j=1}^{i} m_j }{\sum_{j=1}^{i + 1} m_j} \mathbf{p}_{i+1} -  \frac{m_{i+1}}{\sum_{j=1}^{i+1} m_j} \sum_{j=1}^{i} \mathbf{p}_j =\\
        &= \mu_i \left[  \frac{\mathbf{p}_{i+1}}{m_{i+1}} - \frac{\sum_{j=1}^{i} \mathbf{p}_j}{\sum_{j=1}^{i} m_j} \right] \ ,
\end{aligned}
\end{equation}
where $\mu_i$ are the reduced masses defined as
\begin{equation}
    \frac{1}{\mu_i} = \frac{1}{\sum_{j=1}^{i} m_j} + \frac{1}{m_{i+1}} \ .
\end{equation}
The $N$-body system is now described by the linearly independent set of coordinates $(\mathbf{R},\mathbf{r}_{1},...,\mathbf{r}_{N-1})$ in the position space and $(\mathbf{P}, \mathbf{k}_1 ,  ...,\mathbf{k}_{N-1})$ in the momentum space, $\mathbf{R}$ and $\mathbf{P}$ are the center of mass (CM) position and momentum
\begin{equation}
    \mathbf{R} = \frac{1}{M} \ \sum_{i=1}^{N} m_i \ \mathbf{x}_{i} \quad \ ; \quad  \mathbf{P} =  \sum_{i=1}^{N} \mathbf{p}_{i} \ ,
\end{equation}
where $M = \sum_{i=1}^N m_i$ is the total mass.
The transformations from the Cartesian to the Jacobi coordinates $(\mathbf{R},\mathbf{r}_{1}, ...,\mathbf{r}_{N-1}) = T_r \cdot (\mathbf{x}_1,...,\mathbf{x}_N)^{T}$ and for the kinetic momenta $(\mathbf{P}, \mathbf{k}_1 ,  ...,\mathbf{k}_{N-1}) = T_k \cdot (\mathbf{p}_1,...,\mathbf{p}_N)^{T}$ are given by the block matrices:
\begin{equation}
T_r =
    \left(
\begin{array}{ccccc}
 \frac{m_1}{M} &  \frac{m_2}{M} & \frac{m_3}{M} & \cdots & \frac{m_N}{M} \\
 -1 & 1 & 0 & \cdots & 0 \\
 -\frac{m_1}{m_1 + m_2} & -\frac{m_2}{m_1 + m_2} & 1 & \cdots
&  0 \\
 \vdots &  & & & \vdots \\
  0 & 0 & 0 & \cdots &  1 \\
\end{array}
\right)
\end{equation}
and 
\begin{equation}
T_k =
    \left(
\begin{array}{ccccc}
 1 &  1 & 1 & \cdots & 1 \\
 - \frac{m_2}{m_1 + m_2} & \frac{m_1}{m_1 + m_2} & 0 & \cdots & 0 \\
 -\frac{m_3}{m_1 + m_2 + m_3} & -\frac{m_3}{m_1 + m_2 + m_3} & \frac{m_1 + m_2}{m_1 + m_2 + m_3} & \cdots
&  0 \\
 \vdots &  & & & \vdots \\
  0 & 0 & 0 & \cdots &  1 \\
\end{array}
\right) \ .
\end{equation}
Applying the transformation $T_k$ to the Hamoltonian in Eq. \eqref{Hnbody} we have 
\begin{equation}
    \mathcal{H} = \mathcal{H}_{CM} + \sum_{i=1}^{N-1} \mathcal{H}_{i} = \frac{\mathbf{P}^2}{2\ M} + \sum_{i=1}^{N-1}  \frac{\mathbf{k}_i^2}{2 \ \mu_i} \ .
    \label{Hnrel}
\end{equation}

\bibliographystyle{unsrt}      
\bibliography{biblio}   

%
%

\end{document}